\documentstyle[prl,tighten,aps,epsf,amstex,multicol,times]{revtex}
\begin{document}

\draft

\title{The Schmidt Measure as a Tool for Quantifying Multi-Particle Entanglement}

\author{Jens Eisert$^{1,2}$ and Hans J.\ Briegel$^{3}$}
\address{(1) Institut f{\"u}r Physik, Universit{\"a}t Potsdam, 14469
Potsdam,
Germany}
\address{(2) Blackett Laboratory, Imperial College of Science, Technology 
and 
Medicine, London, SW7 2BW, UK}
\address{(3) Sektion Physik, Ludwig-Maximilians-Universit{\"a}t
M{\"u}nchen, 80333 M{\"u}nchen, Germany}
\date{\today}
\maketitle

\begin{abstract}
We present a measure of quantum entanglement which is capable of
quantifying the degree of entanglement of a multi-partite
quantum system. This measure, which is based on a generalization
of the Schmidt rank of a pure state, is
defined on the full state space and is shown to be an entanglement
monotone, that is, it cannot increase under
local quantum operations with classical communication
and under mixing. For a large class of mixed states this measure of 
entanglement can be calculated exactly, and it provides a 
detailed classification of mixed states. 
\end{abstract}


\begin{multicols}{2}
\narrowtext



Quantum entanglement is a still not fully understood
feature of quantum mechanics. 
%
The nature of quantum correlations has 
been a central issue of long-lasting
debates on the interpretation of quantum mechanics.
With the emergence of quantum information theory, however, a more
pragmatic approach to entanglement has become appropriate.
Entanglement, in this context, is typically conceived as a
resource for certain computational and communicational tasks.
Terms like ``entanglement distillation'' -- a process in which
dilute entanglement is transformed into a more suitable
form -- already suggests that emphasis is put on the usefulness of quantum
entanglement.
In the spirit of this development, the
conditions on any functional that is to {\em quantify} the degree
of entanglement  \cite{Bennett1,Quant1,Vidal,Horobound}
have been formulated 
in terms of its behaviour under certain
local quantum
operations that can be implemented by remotely located parties. 

For bi-partite systems
several measures of mixed state entanglement have
been proposed \cite{Bennett1,Quant1,BiPart}, each
supplemented with a different physical interpretation.
In the case of a
multi-partite system 
\cite{Vidal,Multi1,Greenberger,Brun,Duer,Lattice,Classes}
a single number giving the
amount of entanglement is not sufficient: it has been
shown that several ``kinds of entanglement'' have to
be distinguished,
and the quantification of the entanglement
of such multi-partite quantum systems
is essentially an open problem. 

In this paper we introduce a functional -- 
in the following referred to
as Schmidt measure -- that
gives an account on the degree
of entanglement of a multi-partite system with
subsystems of arbitrary dimension.
We will proceed as follows. 
Firstly, the measure of entanglement will be defined for
pure states, employing a natural generalization of the
Schmidt rank for bi-partite systems. 
The definition will then be generalized to the domain
of mixed states, and we will show that this definition
is consistent with conditions every entanglement measure
has to satisfy. 
This entanglement measure provides the tool for a
classification of the entanglement inherent in multi-particle
states. Several partitions \cite{Classes}
of the composite systems in subsystems 
will be considered.
Also, the connection to best separable approximations
as studied in Ref.\ \cite{Lewenstein} will be established.

In Ref.\ \cite{Lattice} a certain class of 
multi-qubit states has been introduced, the
so-called 
$N$-party cluster states $|\phi_N\rangle\langle\phi_N|$.
It has been demonstrated that
the minimal number 
of product terms is given by $2^{\lfloor N/2 \rfloor}$, if
one expands $|\phi_N\rangle$ in
product states of $N$ qubits. 
The need for a 
quantification of the entanglement of
such states was the motivation for the 
investigations of this paper. 
The observation concerning the
minimal number of product states
is related to the findings of
Ref.\ \cite{Duer}, in which
such numbers have first been
considered: it
has been shown that
there are two classes of tripartite entangled pure 
states of three qubits which cannot be transformed
into each other with nonvanishing probability,
the so-called W state \cite{Duer}
and the  GHZ state \cite{Greenberger} 
being representatives. 
One has three and the other one has two product 
states in the minimal decomposition  
in terms of product states. 
This statement is also made stronger 
in that it is pointed out that this minimal number
of product terms can never be increased by means of
invertible local operations. Building upon these observations
one can define an entanglement
monotone on the entire state space, containing the mixed states,
of an arbitrary 
multi-partite system. 


Consider an $N$-partite quantum system
with parties $A_1,...,A_N$
holding quantum systems with dimension $d_1,...,d_N$,
that is, the state space
of the composite system is given by ${\cal S}({\cal H})$, 
where
${\cal H}={\Bbb{C}}^{d_1}\otimes ...\otimes {\Bbb{C}}^{d_N}$.
Any $|\psi\rangle\in {\cal H}$ 
can be written in the form 
\begin{equation}\label{Dec}
	|\psi\rangle=\sum_{i=1}^R
	\alpha_i
	|\psi^{(i)}_{A_1}\rangle \otimes ...\otimes 
	|\psi^{(i)}_{A_N}\rangle,
\end{equation}
where $|\psi^{(i)}_{A_j}\rangle\in{\Bbb{C}}^{d_j}$,
$j=1,...,N$, $\alpha_i\in{\Bbb{C}}$, $i=1,...,R$
with some $R$. Let $r$ be the minimal 
number of product terms $R$ in such a decomposition
of $|\psi\rangle$.
The Schmidt measure 
is now defined as
$P(|\psi\rangle\langle\psi|)=\log_2 r$ \cite{Persistency}.
In case of a bi-partite system with parties $A_1$ and $A_2$ 
the minimal number of product terms $r$ is 
given by the Schmidt rank of the state. The Schmidt
measure could be conceived as a generalization of the Schmidt rank
to multi-partite systems in a similar way as the Schmidt number
applies its concept to mixed states
\cite{Terhal}.

Once $P$ is defined for pure states one
can extend the definition to
the full state space 
in a natural way. This is done by
using a convex roof construction \cite{Bennett1,Uhlmann}:
For a $\rho\in {\cal S}({\cal H})$
let
\begin{equation}\label{Definition}
        P(\rho)=\min \sum_i \lambda_i P(|\psi_i\rangle\langle\psi_i|),
\end{equation}
where the minimum is taken over all possible
convex combinations of the form $\rho=\sum_i \lambda_i 
|\psi_i\rangle\langle\psi_i|$ in terms of pure states
$|\psi_1\rangle\langle\psi_1|,|\psi_2\rangle\langle\psi_2|,...\,$,
with
$0\leq \lambda_i\leq 1$ for all $i$.

In the subsequent paragraph we will show that
the Schmidt measure for mixed states defined in 
Eq.\ (\ref{Definition})
qualifies for being an entanglement monotone 
in the sense of Ref.\ 
\cite{Vidal}( see also Ref.\ \cite{Horobound}). 
That is,   $P$  is a 
proper measure of entanglement  
satisfying the following conditions:
\begin{itemize}
\item[(i)]   $P\geq 0$, and   $P(\rho)=0$ if $\rho$
is fully separable \cite{Fully}.
\item[(ii)]   $P$ is a convex functional, 
\begin{equation}
        P(\lambda \rho_1 + (1-\lambda) \rho_2)\leq
        \lambda P(\rho_1)+(1-\lambda)P(\rho_2)
\end{equation}
for all $\lambda\in[0,1]$ and all $\rho_1,\rho_2\in{\cal S}({\cal H})$.
\item[(iii)]   $P$ is monotone under local generalized
measurements: Let $\rho$ be the initial state, and let
one of the parties perform a 
(partly selective)
local generalized measurement leading to the final 
states $\rho_1,...,\rho_n$ with respective probabilities
  $p_1,...,p_n$. Then 
        $P(\rho)\geq \sum_{i=1}^n p_i P(\rho_i)$.
\end{itemize}
The physical interpretation of condition (i) is
obvious. In condition (ii) part of the distinction between classical
and quantum correlations is incorporated. If one
mixes states in the sense that one dismisses
the classical knowledge of what preparation procedure
has actually been applied, the entanglement
must not increase.
Condition (iii) ensures that 
the expected entanglement 
does not grow when a selective generalized
measurement is performed.
In particular, condition (iii) leads to an
invariance under local unitary operations, that is,
  $P(U\rho U^\dagger)=P(\rho)$ for all $\rho\in{\cal S}({\cal H})$
and all local unitary operators $U$.
An implication of (ii) and (iii) together is that
  $P( \Lambda(\rho))\leq P(\rho)$ for all $\rho\in{\cal S}({
\cal H})$
and all completely positive trace-preserving maps 
$\Lambda:{\cal S}({\cal H})\longrightarrow{\cal S}({\cal H})$
corresponding to a local quantum operation.
These conditions (i) -- (iii) that formalize the intuitive
properties of an entanglement measure 
go back to Refs.\ \cite{Bennett1} and
\cite{Quant1}.\\

\noindent {\bf Proposition. --}   $P$ is an entanglement
monotone fulfilling 
the conditions (i) -- (iii). 

{\it Proof:}\/ Condition (i) follows immediately 
from
the definition. Due to the convex roof construction
  $P$ is also a convex functional: 
let $\rho_1$ and 
$\rho_2$ be states from ${\cal S}({\cal H})$, and let
$\rho_1=\sum_j \mu_j |\phi_j\rangle\langle\phi_j|$
and $\rho_2=\sum_j \eta_k |\varphi_k\rangle\langle\varphi_k|$
be the two decompositions
for which the respective minima
in Eq.\ (\ref{Definition}) are attained.
Then 
$	\sum_j \lambda \mu_j |\phi_j\rangle\langle\phi_j|
	+ \sum_k (1-\lambda) 
	\eta_k |\varphi_k\rangle\langle\varphi_k|$
is a valid decomposition of $\rho\equiv
\lambda \rho_1 + (1-\lambda) \rho_2$, but it is 
not necessarily the optimal one. Hence,
  $P(\lambda \rho_1 + (1-\lambda) \rho_2)\leq
        \lambda P(\rho_1)+(1-\lambda)P(\rho_2)$.

To see that   $P$ 
satisfies also condition
(iii) we proceed in several steps. The local measurement
can be assumed to be performed by party $A_1$. 
Any final state $\rho_j$, $j=1,2,...$, in a 
local generalized measurement can be 
represented with the help of
Kraus operators $E_{ij}$, $i,j=1,2,...$, as
\begin{equation}
        \rho_j=\frac{\sum_i {E_{ij}}\rho {E_{ij}}^\dagger}{p_j
}
\end{equation}
with   $p_j={\rm tr}[\sum_i {E_{ij}}\rho {E_{ij}}^\dagger]$,
$\sum_{ij}E_{ij}^\dagger E_{ij}=I$.
As the sum over $i$ corresponds to a 
mixing which -- due to the convexity property --
reduces the Schmidt measure, it 
suffices to consider final states of the form
$\rho_j={E_{j}}\rho {E_j}^\dagger/p_j$
with   $p_j={\rm tr}[{E_{j}}\rho {E_j}^\dagger]$.
The trace-preserving property of
the quantum operation reads as
$\sum_{j} E_j^\dagger  {E_{j}}=I$.
For any pure state  $|\psi\rangle\langle\psi|\in{\cal S}({\cal H})$
\begin{equation}\label{Crucial}
        P({E_{j}}|\psi\rangle\langle\psi| {E_j}^\dagger/
        {\rm tr}[
        {E_{j}}|\psi\rangle\langle\psi| {E_j}^\dagger
        ])\leq P(|\psi\rangle\langle\psi|)
\end{equation}
for all $j=1,2,...\,$. This can be seen as follows.
Let $|\psi\rangle=\sum_{i=1}^r|\psi^{(i)}_{A_1}\rangle
\otimes ...\otimes 
|\psi^{(i)}_{A_N}\rangle$ 
be the decomposition of $|\psi\rangle$
into products as in Eq.\ (\ref{Dec}) 
with the minimal number of terms $r$. Then $E_j$ is
either invertible, and then $E_j |\psi\rangle$ has the same
minimal number of product terms $r'=r$, or it is not invertible, such 
that $r'\leq r$.
Moreover, 
if Eq. (\ref{Crucial}) holds for pure states $|\psi\rangle
\langle\psi|$,
it is also valid for arbitrary states $\rho\in {\cal S}({\cal 
H})$: Let
$\rho=\sum_k \lambda_k |\psi_k\rangle\langle\psi_k|$ be the
optimal decomposition of $\rho$ belonging to the minimum in Eq.\ 
(\ref{Definition}), then 
\begin{eqnarray}
      P(\rho)&=&\sum_k \lambda_k P(|\psi_k\rangle\langle\psi_k|)
        \nonumber\\
        &\geq& \sum_k \lambda_k P\bigl(
        E_{j}|\psi_k\rangle\langle\psi_k|E_j^\dagger/
        {\rm tr}[
        E_{j}|\psi_k\rangle\langle\psi_k|
	E_j^\dagger]\bigr)\nonumber\\
        &\geq& P\bigl(E_{j}\rho E_j^\dagger/
        {\rm tr}[
        E_{j} \rho E_{j}^\dagger
        ]\bigr)
\end{eqnarray}
for all $j=1,2,...\,$.
The statement of condition (iii) follows from the fact that
$       \sum_j p_j P( E_{j}\rho  E_j^\dagger/{\rm tr}[
         E_{j}\rho  E_{j}^\dagger
        ])
        \leq \sum_j p_j P(\rho)=P(\rho)$.
\hfill\fbox\\
\bigskip


The Schmidt measure can indeed be used 
as a functional appropriately
quantifying the entanglement of a given state of a 
$N$-partite quantum system. 
It cannot be increased on average
under LOCC
and it distinguishes between classical and
quantum correlations \cite{Remark}. 
The Schmidt measure 
is normalized, in that
	$P(|\psi\rangle\langle\psi|)=1 $
for all states 
$|\psi\rangle\langle\psi|$ of the form
$|\psi\rangle=(|0_1...0_N\rangle+|1_1...1_N\rangle)/\sqrt{2}$.
The functional is also (fully) additive on pure states:
If the parties $A_1,...,A_N$ share two $N$-partite
quantum systems in the states $|\psi_1\rangle\langle\psi_1|$
and $|\psi_2\rangle\langle\psi_2|$, respectively, then
  $P(|\psi_1\rangle\langle\psi_1|\otimes
|\psi_2\rangle\langle\psi_2|)=P(|\psi_1\rangle\langle\psi_1|)
+P(|\psi_2\rangle\langle\psi_2|)$.
In particular,   $P(|\psi\rangle\langle\psi|^{\otimes n})=
n P(|\psi\rangle\langle\psi|)$
for $n$ copies of the same state $|\psi\rangle\langle\psi|$.
Let
$\rho\in {\cal S}({\cal H})$ be a fully 
separable state \cite{Fully}, then the stronger
statement
  $P( \rho^{\otimes n}\otimes
|\psi\rangle\langle\psi|^{\otimes n})=
n P(|\psi\rangle\langle\psi|)$ is also
true \cite{Subadd}.


Although the
Schmidt measure of a mixed state 
is defined via a minimization over all possible
realizations of the state,
it can be calculated exactly for a quite large
class of states. This is mainly due to 
the fact
that it is a coarse grained measure.
All terms that appear in Eq.\ (\ref{Definition})
are logarithms of natural numbers weighted with respective
probabilities. 

This fact allows for a 
detailed classification of multi-particle
entangled states. 
The following investigations
will be restricted to the multi-qubit case, meaning 
that ${\cal H}=({{\Bbb{C}}^2})^{\otimes N}$.
Clearly, in a multi-partite setting a single
number for the entanglement of the whole system
is not sufficient to account 
for the possible entanglement structures.
As in Ref.\ \cite{Classes} we consider arbitrary partitions
of the $N$-partite system
with parties $A_1,...,A_N$
into $k$ parts, $k=2,...,N$. A division of
the original system
into  $k$ parts will be called a 
$k$-split. 
These parts are
taken to be a system on their own with a certain
higher dimension when   $P$
is evaluated. In notation such a 
division will be marked with brackets.
For a, say, 
three-party system the 3-split $A_1 A_2 A_3$
and the three 2-splits $(A_1 A_2) A_3$, $(A_2 A_3) A_1$,
and $(A_3 A_1) A_2$
are possible. In this classification the 
Schmidt measure will also be furnished with 
an index indicating the respective split.
For each $k$-split the Schmidt measure does not increase 
on average in the course of
a LOCC operation. This does not imply, however,  that
the Schmidt measure with respect to the reduced state
of some of the parties must not increase 
on average under such operations.

{\it Pure states. --\/} 
For pure states the definition given in Eq.\ 
(\ref{Dec})
can be easily applied.  
Take, e.g., as in Ref.\ \cite{Duer} 
the three party W-state with state vector 
\begin{equation}
|\text{W}\rangle=
(|100\rangle+|010\rangle+|001\rangle)/\sqrt{3}.
\end{equation}
  $P_{A_1 A_2 A_3}(|\text{W}\rangle\langle \text{W}|)=\log_2 3$, 
while the three party GHZ state,
$|\text{GHZ}\rangle=(|000\rangle+|111\rangle)/\sqrt{2}$,
gets the value   $P_{A_1 A_2 A_3}(|\text{GHZ}\rangle\langle 
\text{GHZ}|)=1$. 
Hence, in the 3-split
  $P$ distinguishes the GHZ state, the
W state and product states (value 0) \cite{Tangle}. 
Other splits with $k=2$ reveal further information and give rise to
the full 
classification. Table \ref{Table1}
shows the values of the Schmidt measure with respect to
all possible splits for some pure states of a
four-partite system.

\noindent \begin{minipage}{1.\columnwidth}
\begin{table}
\caption{Values of $P$ for 
some four-qubit pure states:
the four-party GHZ state with state vector
$(|0000\rangle+|1111\rangle)/\sqrt{2}$, the
generalized W state [9] corresponding to 
$(|0001\rangle +|0010\rangle +|0100\rangle+|1000\rangle
)/2$, the cluster state of Ref.\ [10] ($|\phi_4\rangle=
(|0000\rangle+|0011\rangle+|1100\rangle-|1111\rangle)/2$),
and the product of two maximally entangled states of
two qubits
($|\phi^+\rangle |\phi^+\rangle=
(|00\rangle+|11\rangle)\otimes (|00\rangle+|11\rangle)/2$).
The values of $P$ 
associated with the remaining splits 
can be obtained from the permutation 
symmetry of the states.}

\begin{center}
\begin{tabular}{|c||c|c|c|c|}
 & GHZ & W &  $|\phi_4\rangle$ & $|\phi^+\rangle |\phi^+\rangle$ \\
\hline
$A_1 A_2 A_3 A_4$     & 1 & 2 & $2$ & $2$\\
$(A_1 A_2) A_3 A_4$   & 1 & $\log_2 3$ & $1$ & 1\\
$(A_1 A_2) (A_3 A_4)$ & 1 & $1$ & $1$ & 0\\
$(A_1 A_3) (A_2 A_4)$ & 1 & $1$ & $2$ & $2$\\
$(A_1 A_2 A_3) A_4$   & 1 & 1 & 1 & 1\\
\end{tabular}
\end{center}
\label{Table1}
\end{table}
\end{minipage}

{\it Mixed states. --\/} 
In mixed quantum states both 
quantum entanglement and
classical correlations 
may be present.
In order to calculate the Schmidt measure of a mixed state
a minimization over decompositions of the state
is required. Upper bounds of   $P$
follow immediately from the definition: If
$\rho=\sum\eta_i |\psi_i\rangle\langle\psi_i|$ is
any  not necessarily optimal decomposition of
a state $\rho \in{\cal S}({\cal H})$, then 
$\sum_i\eta_i P(|\psi_i\rangle\langle\psi_i|)$ is
an upper bound of   $P(\rho)$.
For many states   $P$ can however be fully
evaluated. Consider, e.g., two parties $A_1$ and $A_2$
sharing two qubits in the Werner state \cite{Werner}
\begin{equation}
        \rho_{\text{W}}(\lambda)
        =\lambda |\psi^-\rangle \langle \psi^-|+(1-\lambda) 
        I/4,
\end{equation}
with $|\psi^-\rangle=
(|01\rangle-|10\rangle)/\sqrt{2}$, $0\leq \lambda\leq 1$.
As all pure states in the range of $\rho_{\text{W}}(\lambda)$ 
have Schmidt measure 0 or 1, one has to identify in any
decomposition $\rho_{\text{W}}(\lambda)=
\sum_i
\eta_i |\psi_i\rangle\langle\psi_i|$
the terms with Schmidt measure 0 (product states) or 1
(entangled states). Hence, the 
Schmidt measure is given by   $P(\rho_{\text{W}}(\lambda))=1-s$,
where $s$ is the weight of the separable state that
can be maximally subtracted from $\rho_{\text{W}}(\lambda)$
while maintaining the semi-positivity of the state,
and therefore \cite{Lewenstein},
\begin{equation}
        P(\rho_{\text{W}}(\lambda))=\left\{
        \begin{array}{ll}
        \frac{3}{2}\lambda -\frac{1}{2}, & {\text{
        for $1/3<\lambda\leq 1$,}}\\
        0,& {\text{for $0\leq \lambda\leq 1/3$.}}\\
        \end{array}
        \right.
\end{equation}
In other words,   $P$ is just given by
the weight of the inseparable state in the best
separable decomposition in the sense of Ref.\ \cite{Lewenstein}.
As follows from the observation 
that the Schmidt rank of 
all states in the range of any state from 
${\cal S}({\Bbb{C}}^2\otimes {\Bbb{C}}^2)$ is smaller
or equal to 2,
this statement holds for all mixed states of two parties
holding qubits. 
As a corollary we find that
the parameter of the best separable decomposition
in $2\times 2$ systems is an entanglement monotone.

For more than two parties several different entanglement
classes can be distinguished. 
The Schmidt measure is
again defined for all possible $k$-splits, $k=2,...,N$.
If the $N$-partite system is separable with respect
to a particular $k$-split, the value of the
corresponding Schmidt measure is 0.
Since not only condition (i) of the conditions of 
the entanglement monotone is satisfied, but also
the stronger statement that $P(\rho)=0$ iff
$\rho$ is fully separable, the Schmidt measure
with respect to a certain split gives an account on 
the separability of the state taking this split.
For three qubits, e.g., the classes of so-called
one-qubit biseparable states,
two-qubit  biseparable states,
three-qubit  biseparable states, and
fully separable states \cite{Classes}
can be distinguished.
However, the Schmidt measure reveals more structure,
since the entanglement -- if present -- is also quantified. 
As an example, consider the state
\begin{eqnarray}
        \rho(\lambda,\mu)&=&
        \lambda
        |\phi^+_{A_1 A_2 }\rangle|0_{A_3}\rangle \langle 0_{A_3}|
        \langle \phi^+_{A_1 A_2 }|\nonumber \\
        &+&
        \mu|\phi^+_{A_2 A_3 }\rangle|0_{A_1}\rangle \langle 0_{A_1}|
        \langle \phi^+_{A_2 A_3 }|\nonumber \\
        &+&
        (1-\lambda-\mu)
        |\phi^+_{A_3 A_1 }\rangle|0_{A_2}\rangle \langle 0_{A_2}|
        \langle \phi^+_{A_3 A_1 }|,
\end{eqnarray}
$|\phi^+\rangle=
(|00\rangle+|11\rangle)/\sqrt{2}$,
$0\leq \lambda,\mu\leq1$. For $\lambda=\mu=1/3$ this state
reduces to the three-party molecule state $\rho_{\text{M}}$
studied in Ref.\
\cite{Molecules}. The Schmidt measure
  $P_{A_1 A_2 A_3}(\rho_{\text{M}})=1$
is equal to the Schmidt measure 
of the state where $A_1$ and $A_2$
hold a $|\phi^+\rangle\langle\phi^+|$ 
state and $A_3$ is in the state $|0\rangle\langle0|$,
as the mere classical ignorance of which parties 
are actually holding 
the Bell state cannot increase the amount of entanglement
(compare also Ref.\ \cite{Amnesia}). 
In Table \ref{Table2} the values of the Schmidt measures
of all splits of the states $\rho(\lambda,\mu)$,
$\rho_{\text{M}}$, and
$        \rho_{\text{G}}(\lambda)=\lambda
        |\text{GHZ}\rangle\langle \text{GHZ}|
        +(1-\lambda) |000\rangle \langle 000|$,
$0\leq \lambda\leq1$, are shown.

\noindent 
\begin{minipage}{1.\columnwidth}
\begin{table}
\caption{The Schmidt measure   $P$
for some mixed quantum states. 
}
\begin{center}
\begin{tabular}{|c||c|c|c|}
 & $\rho_{\text{G}}(\lambda)$ & $\rho_{\text{M}}$ & $\rho(\lambda,\mu)$ \\
\hline
$A_1 A_2 A_3$      & $\lambda$ & 1 & 1 \\
$(A_1 A_2 ) A_3 $  & $\lambda$  & $2/3$ & $1-\lambda$ \\
$(A_1 A_3) A_2 $   & $\lambda$  & $2/3$ & $\lambda+\mu$ \\
$(A_2 A_3) A_1 $   & $\lambda$  & $2/3$ & $1-\mu$ \\
\end{tabular}
\end{center}
\label{Table2}
\end{table}
\end{minipage}


To summarize, we have proposed a new functional
which quantifies the entanglement
of quantum systems shared by $N$ parties in 
possibly mixed states. 
It has the property
to be an entanglement monotone, implying that it
does increase on average under LOCC
operations. Surprisingly, it turns out to be  
analytically computable for a large 
class of states, and it leads to a detailed
classification of states. It is the hope
that this measure of entanglement 
provides a helpful pragmatic 
tool for investigating the
rich structure of multi-particle entanglement.


We would like to acknowledge fruitful
discussions with  M.B.\ Plenio,  W.\ D{\"u}r, G.\ Vidal,
and
D.\ Bru{\ss} at the Benasque Center for Science (BCS).
One of us (HJB) enjoyed interesting discussions with L. Hardy and
C. Simon during an earlier visit at Oxford University.
We would also like to thank M.B.\ Plenio for critically reading the manuscript.
This work has been supported in part by the DFG 
and the European Union
(IST-1999-11053, 
IST-1999-13021, 
IST-1999-11055).

\end{multicols}


\begin{thebibliography}{99}
\bibitem{Persistency}
	For a number of states one can show 
	that the Schmidt measure coincides with the
	persistency of entanglement 
	introduced in Ref.\ \cite{Lattice}. Examples
	for such states are the cluster states of $N$ qubits
	\cite{Lattice} (see also Table \ref{Table1}).
\bibitem{Fully}
	$\rho\in{\cal S}({\cal H})$ is fully separable if
	it can be 
	written in the form 
	$
	\rho=\sum_i p_i
	|\psi^{(i)}_{A_1}\rangle\langle \psi^{(i)}_{A_1}|\otimes
	... \otimes
	|\psi^{(i)}_{A_N}\rangle\langle \psi^{(i)}_{A_N}|
	$
	with $|\psi^{(i)}_{A_j}\rangle
	\in{\Bbb{C}}^{d_j}$ for all $j=1,...,N$
	and   $p_i>0$, $i=1,2,...\,$.
\bibitem{Remark}
        Among the postulates for an entanglement measure
        in the strict sense of Ref.\ \cite{Horobound}
        $P$ 
        does not satisfy a
        continuity criterion. In particular, it
	is not continuous in a region close to 
	$n$-fold
	products $|\psi\rangle\langle\psi|^{\otimes n}$
	of pure states. Therefore, $P$ does not
	have to coincide with the von-Neumann
	entropy of
	one subsystem for a bi-partite system in a pure state.
\bibitem{Subadd}
        For general mixed states $P$ 
        is subadditive,
          $P(\rho\otimes \sigma )\leq P(\rho)+P(\sigma)$ 
        for all $\rho,\sigma
       \in{\cal S}({\cal H})$, where $\rho$ and $\sigma$
        are states of quantum systems
	held by the same parties.
\bibitem{Tangle}
        The 3-tangle, proposed in Ref.\ \cite{Wootters},
        is also an entanglement monotone (see Ref.\ \cite{Duer}),
        and it can distinguish the W from the GHZ
        state. However, it is defined only for
        three qubits, and it gives the same value for
        the W state and for product states.
\end{thebibliography}
\end{document}